# BER Performance of Photon Counting PPM vs. DPSK for Satellite Communications

*Daniel A. Paulson*

Graduate Student

University of Maryland, College Park

Department of Electrical and Computer Engineering

This work was completed in December, 2015

**Abstract** – Expressions for the BER of *M*-ary PPM & biphase DPSK modulations in the presence of noise are derived using analytical, statistical methods. The PPM expression is verified via Poisson statistics based simulation. BER expressions are then applied to a representative set of receiving telescope & sky spectral radiance parameters in order to assess performance of PPM & DPSK relative to one another. Finally, efficiency & additional considerations are discussed.

## I. INTRODUCTION

The National Aeronautics and Space Administration (NASA) current is augmenting its communications technology capabilities to include free space optical communications system for near earth & deep Space applications. The Lunar Laser communications demonstration has already demonstrated a 622 Mbps optical pulse position modulation (PPM) link in deep space [1], and now NASA is investing in the Laser Communications Relay Demonstration (LCRD) project, which will add a direct detection biphase differential phase shift keying (DSPK) modulation capability [2]. Both modulation types are slated to support the same set of data rates, from 2 Mbps to 1.244 Gbps at optical C-band, so it is a natural question to ask: Which modulation type is expected to perform better in the space communications arena? In this work, we will derive expressions for expect bit error rates (BER) in the presence of noise for the competing technologies, & evaluate performance using a representative example receiver telescope scenario.

## II. ABSOLUTE LIMITS FOR PPM SYMBOL ERROR RATE (SER)

Prior to reviewing the performance in the presence of noise, it is a useful exercise to discuss limits to the bit error rate due to the randomness of photon arrival times at the detector. In the case of photon counting this is a time-varying Poisson process [3]. In the absence of noise, a theoretical detectors probability of detecting $n$ photons in a given period is:

$$p(n) = \frac{N_{ph}{}^n e^{-N_{ph}}}{n!}$$

Where $N_{ph}$ is the mean number of photons per period, also often represented as $N_{ph} = \lambda(t)T$ with $\lambda(t)$ the intensity function & $T$ the period. Because it is relevant as background for PPM analysis, investigating the case of On-Off Keying (OOK), without noise there would be no probability of detecting a signal when none was sent, & probability $p(0) = \frac{N_{ph}{}^0 e^{-N_{ph}}}{0!} = e^{-N_{ph}}$ of receiving no photons when the transmitter was attempting to send a signal. The total probability of error, assuming *ons* & *offs* are sent with equal likelihood, is therefore the sum of the prior 2 probabilities, or $\frac{1}{2}e^{-N_{ph}}$ where we explicitly define $N_{ph}$ as the average number of photons for an *on* period. If we define an $N'_{ph}$ as the average number of photons transmitted during both *on* and *off* periods, the probability of error becomes $\frac{1}{2}e^{-2N'_{ph}}$.

For 2-ary PPM, where the probability of *ons* & *offs* will be strictly enforced as ½ for any bit sequence, the probability of not detecting an *on* is again $e^{-N_{ph}}$ & the probability of not detecting an *off* is 0. However, in the case that the *on* for a given symbol period is not detected, the receiver must decide to assign a 1 or a 0 to the symbol, with the associated likelihood of error for this case being ½ (totally random). The probability of error is therefore exactly the same as for OOK:

$$PE_{2-ary\ PPM} = \frac{1}{2}e^{-N_{ph}} = \frac{1}{2}e^{-2N'_{ph}}$$

Since for *M-ary* PPM, the probability of not detecting any photons during an on period is still $e^{-N_{ph}} = e^{-MN'_{ph}}$. Should no photons be detected, the probability of a symbol error is $\frac{M-1}{M}$, therefore:

$$PE_{M-ary\ PPM} = \frac{M-1}{M}e^{-N_{ph}} = \frac{M-1}{M}e^{-MN'_{ph}}$$

For uncoded data, assuming $M$ a power of 2, when an error occurs the probability of each bit represented by the symbol being in error is ½. Therefore, the uncoded bit error rate (BER) would be $BER_{uncoded} = \frac{1}{2}e^{-N_{ph}} = \frac{1}{2}e^{-MN'_{ph}}$. This is a bit counterintuitive, because it would appear in this noise-free environment that we can increased the efficiency of our system by increasing $M$ arbitrarily, however note that when a symbol error does occur a whole $log_2(M)$ bits in a row will be entirely random. This could lead to long outages in useable signal & if more uncorrectable frames in general if block coding is employed.

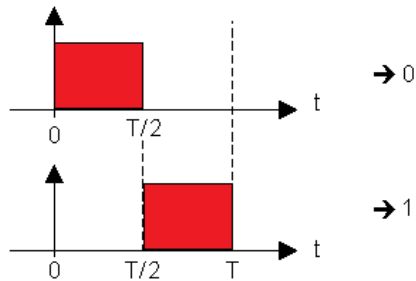

*Fig. 1 - Example signal for 2-ary PPM.*

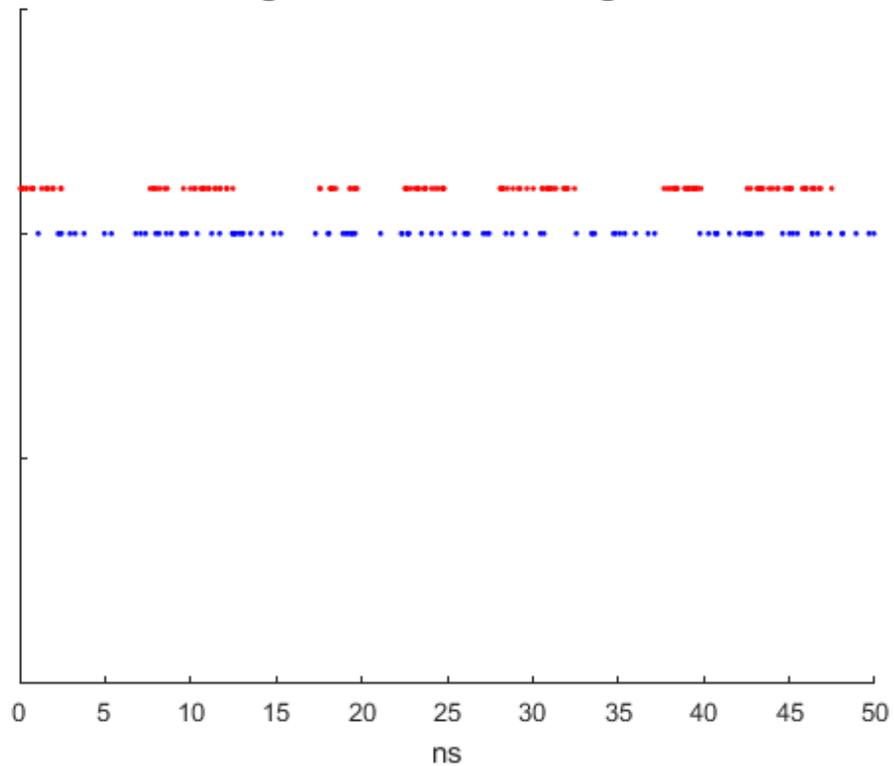

*Fig. 2* - **Illustration of the semi-random nature of photon counts & arrival times for 200 Mbps 2-ary PPM signal (red) along with noise photons (blue). Note that for most optical applications, the PPM photons are likely to be pulsed, with less spread per slot.**

III. THEORETICAL PERFORMANCE OF 2-ary PPM IN THE PRESENCE OF NOISE

In the presence of background noise photons, the probability of a symbol error is given by the sum of the probabilities of an *off* time slot having more photon counts that an *on* time slot & ½ of the probability an *on* and *off* slot having the same amount of counts. Utilizing Poisson statistics, for 2-ary PPM the probability of a symbol error is given by:

$$PE_{2-ary\ PPM} = \sum_{k_1=0}^{\infty} \frac{(K_s + K_b)^{k_1}}{k_1!} e^{-(K_s+K_b)} \left( \sum_{k_2=k_1+1}^{\infty} \frac{K_b^{k_2}}{k_2!} e^{-K_b} + \frac{1}{2} \frac{K_b^{k_1}}{k_1!} e^{-K_b} \right)$$

$$= \sum_{k_1=0}^{\infty} \frac{(K_s + K_b)^{k_1}}{k_1!} e^{-(K_s+K_b)} \sum_{k_2=k_1}^{\infty} \frac{K_b^{k_2}}{k_2!} e^{-K_b} \left( 1 - \frac{\delta_{k_1 k_2}}{2} \right)$$

$$= \sum_{k_1=0}^{\infty} \frac{(K_s + K_b)^{k_1}}{k_1!} e^{-(K_s+K_b)} \left[ \left( \sum_{k_2=k_1}^{\infty} \frac{K_b^{k_2}}{k_2!} e^{-K_b} \right) - \frac{1}{2} \frac{K_b^{k_1}}{k_1!} e^{-K_b} \right]$$

$$= \left( \sum_{k_1=0}^{\infty} \frac{(K_s + K_b)^{k_1}}{k_1!} e^{-(K_s+K_b)} \sum_{k_2=k_1}^{\infty} \frac{K_b^{k_2}}{k_2!} e^{-K_b} \right) - \frac{1}{2} \left( \sum_{j=0}^{\infty} \frac{(K_s + K_b)^j}{j!} e^{-(K_s+K_b)} \frac{K_b^j}{j!} e^{-K_b} \right)$$

Where $K_s$ is the expected number of signal counts per *on* time slot, $K_b$ is the expected number of background counts per time slot, & $\delta$ is the Kronecker delta function. It is easily verified that the above equations are equal to the expression derived as an absolute limit without the presence of background photons. This expression can be written in another form, utilizing the Marcum Q function [4], and the definition:

$$Q(a,b) \triangleq \int_b^{\infty} x\, e^{\frac{-(a^2+x^2)}{2}} I_0(ax)\, dx$$

Where $I_0$ is the $0^{th}$ modified Bessel function of the $1^{st}$ kind. Using the identity:

$$Q(a,b) = \sum_{k_1=0}^{\infty} \frac{\left(\frac{b^2}{2}\right)^{k_1}}{k_1!} e^{\frac{-b^2}{2}} \sum_{k_2=k_1}^{\infty} \frac{\left(\frac{a^2}{2}\right)^{k_2}}{k_2!} e^{\frac{-a^2}{2}}$$

Is identity brings our expression for the BER to:

$$PE_{2-ary\ PPM} = Q\left(\sqrt{2K_b}, \sqrt{2(K_b + K_b)}\right) - \frac{1}{2}\left(\sum_{j=0}^{\infty} \frac{(K_s + K_b)^j}{j!} e^{-(K_s+K_b)} \frac{K_b{}^j}{j!} e^{-K_b}\right)$$

The 2$^{nd}$ term can be simplified using the following relation [5]:

$$I_0(z) = \sum_{k=0}^{\infty} \frac{\left(\frac{z}{2}\right)^{2k}}{(k!)^2}$$

Therefore:

$$PE_{2-ary\ PPM} = Q\left(\sqrt{2K_b}, \sqrt{2(K_s + K_b)}\right) - \frac{1}{2} e^{-(K_s+2K_b)} I_0\left(2\sqrt{(K_s + K_b)K_b}\right)$$

### IV.   THEORETICAL PERFORMANCE OF *M*-ary PPM IN THE PRESENCE OF NOISE

Deriving a closed form, general expression for *M*-ary PPM in the presence of background counts is more difficult, due to the problem of handling tie-breaking. For instance, for $M = 8$ there exists a chance of an 8-way tie in which case the probability of symbol for the tie would be $\frac{7}{8}$, in the case of a 7-way tie in which case the probability of symbol for the tie would be $\frac{6}{7}$, and so on. Many papers discussing the error rate of PPM using avalanche photodiodes are able to avoid this issue by assuming that the noise is Gaussian as opposed to a countable number of photons [6] [7]. A simple upper bound on the SER, is provided by Hughes [8]:

$$PE_{M-ary\ PPM}(K_s, K_b) \leq 1 - \left(1 - PE_{M-ary\ PPM}(K_s, K_b)\right)^M$$

$$= 1 - \left(Q\left(\sqrt{2K_b}, \sqrt{2(K_s + K_b)}\right) - \frac{1}{2} e^{-(K_s+2K_b)} I_0\left(2\sqrt{(K_s + K_b)K_b}\right)\right)^M$$

If one is willing to consider numerically calculated solutions, we can write the symbol error rate in a form which considers both the probability $1$ to $M - 1$ of the *off* slots contained more photons than the *on* slot & in the case that there is an m way tie (m equal 2 through *M*). There are two conditions that can cause a symbol error, which we must consider:

1) 1 through $M - 1$ *off* slots have greater photon counts than the *on* slot

2) We have 1 through $M-1$ *off* slots tied with the *on* slot for most photon counts in the symbol period, or stated another way, a 2 through $M$-way tie with the *on* slot included.

Scenario 1) will cause a symbol error with 100% probability, but this will not be the case for scenario 2).

The probability of 1), that $m = 1$ through $M-1$ of the *off* slots have greater counts than the *on* slot count, $k_1$, for a given $k_1$ is:

$$\binom{M-1}{m} \left( \sum_{k_2=k_1+1}^{\infty} \frac{K_b^{k_2}}{k_2!} e^{-K_b} \right)^m \left( \sum_{k_3=0}^{k_1-1} \frac{K_b^{k_3}}{k_3!} e^{-K_b} \right)^{M-m-1}$$

Where we make use of the binomial coefficient, $\binom{M-1}{m} = \frac{(M-1)!}{m!(M-1-m)!}$. The probability of condition 2), an $(m+1)$-way tie for a given $k_1$, is:

$$\binom{M-1}{m} \left( \frac{K_b^{k_1}}{k_1!} e^{-K_b} \right)^m \left( \sum_{k_3=0}^{k_1-1} \frac{K_b^{k_3}}{k_3!} e^{-K_b} \right)^{M-m-1}$$

If in the face of a tie, we assume the symbol is chosen randomly from the tied slots, the probability of error is given an $(m+1)$-way tie is $\frac{m}{m+1}$.

Bearing in mind the probability of a given *on* slot photon count, $k_1$, we consider the probability of error for a tie, & sum over the valid values of $m$ to give our final expression for the probability of symbol error for $M$-ary PPM:

$$PE_{M-ary\ PPM} = \sum_{k_1=0}^{\infty} \frac{(K_s + K_b)^{k_1}}{k_1!} e^{-(K_s+K_b)} \sum_{m=1}^{M-1} \binom{M-1}{m} \left( \left( \sum_{k_2=k_1+1}^{\infty} \frac{K_b^{k_2}}{k_2!} e^{-K_b} \right)^m \right.$$
$$\left. + \frac{m}{m+1} \left( \frac{K_b^{k_1}}{k_1!} e^{-K_b} \right)^m \right) \left( \sum_{k_3=0}^{k_1-1} \frac{K_b^{k_3}}{k_3!} e^{-K_b} \right)^{M-m-1}$$

I was unable to found a separate confirmation that this probability of error formula is correct in the literature. For this reason, I have verified my solution is correct via simulation of Poisson signal & noise processes. Given a signal photon count of 20 photons per symbol, and background counts of 4 photons per symbol, running a simulation of M = 2 through 128-ary PPM testing each applicable power of 2 for 500 Megasymbols, we see the simulation results match the derived expression closely. Because one cannot sum to infinity in a finite amount of time, the summations over $k_1$ & $k_2$ for the theoretical results were capped at 10,000.

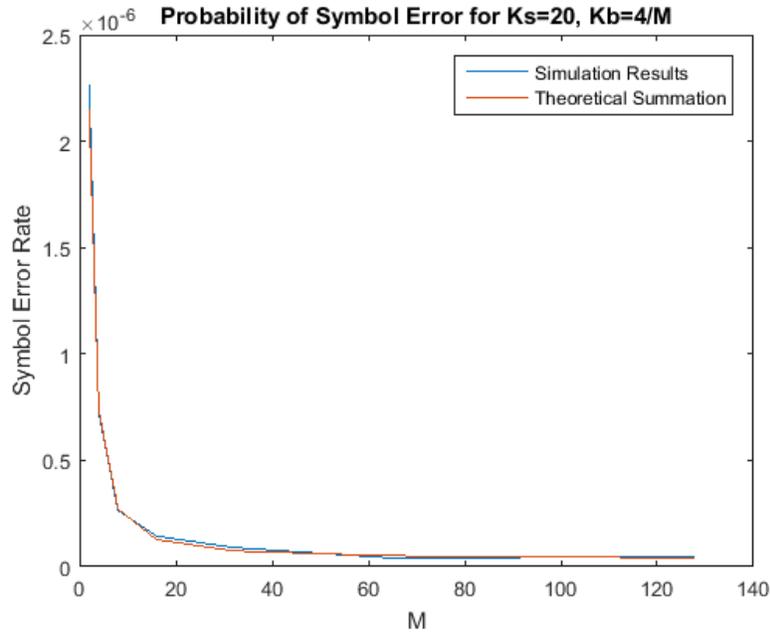

*Fig. 3 - PPM linear scale[1] theoretical versus simulation results for 500 Megasymbols.*

Should a symbol error occur, one of $M-1$ slots has been selected with equal probability (assuming random data) of $\frac{PE_{M-ary\ PPM}}{M-1}$, or for $b = log_2 M$ bits per symbol, $\frac{PE_{M-ary\ PPM}}{2^b-1}$ [9]. The number of ways in which $n$ of the $b$ bits can be in error is $\binom{b}{n}$. Since:

---

[1] For a logarithmic scale plot, see the next figure, where the same data is plotted alongside BER plots on logarithmic x & y axis scales.

$$\left\langle \frac{bit\ errors}{symbol} \right\rangle = \sum_{n=1}^{b} n \binom{b}{n} \frac{PE_{M-ary\ PPM}}{2^b - 1} = \sum_{n=1}^{b} n \frac{b!}{n!\,(b-n)!} \frac{PE_{M-ary\ PPM}}{2^b - 1}$$

$$= \sum_{n=1}^{b} \frac{b!}{(n-1)!\,(b-n)!} \frac{PE_{M-ary\ PPM}}{2^b - 1} = \sum_{n=0}^{b-1} \frac{b!}{(n)!\,(b-n-1)!} \frac{PE_{M-ary\ PPM}}{2^b - 1}$$

$$= b \cdot \sum_{n=0}^{b-1} \frac{(b-1)!}{(n)!\,(b-1-n)!} \frac{PE_{M-ary\ PPM}}{2^b - 1} = b \cdot \sum_{n=0}^{b-1} \binom{b-1}{n} \frac{PE_{M-ary\ PPM}}{2^b - 1}$$

$$= b \cdot 2^{b-1} \cdot \frac{PE_{M-ary\ PPM}}{2^b - 1} = b \cdot \frac{2^{b-1}}{2^b - 1} \cdot PE_{M-ary\ PPM}$$

Where we have used the identity $2^x = \sum_{y=0}^{x} \binom{x}{y}$, the probability of a bit error is the above expression divided by the number of bits per symbol, b. Dividing by $b$ & substituting $M = 2^b$, $\frac{M}{2} = 2^{b-1}$, & using the previously derived expression for $PE_{M-ary\ PPM}$, we arrive at a general expression for BER of $M$-ary PPM:

$$BER_{M-PPM} = \frac{M}{2(M-1)} \sum_{k_1=0}^{\infty} \frac{(K_s + K_b)^{k_1}}{k_1!} e^{-(K_s+K_b)} \sum_{m=1}^{M-1} \binom{M-1}{m} \left( \left( \sum_{k_2=k_1+1}^{\infty} \frac{K_b^{k_2}}{k_2!} e^{-K_b} \right)^m \right.$$

$$\left. + \frac{m}{m+1} \left( \frac{K_b^{k_1}}{k_1!} e^{-K_b} \right)^m \right) \left( \sum_{k_3=0}^{k_1-1} \frac{K_b^{k_3}}{k_3!} e^{-K_b} \right)^{M-m-1}$$

Which we confirm via 500 Mega-symbol simulation below. The slight error at the right hand tail of the Figure 4 (below) plots can bit attributed to the low number of errors for higher order PPM at this level of noise. For instance, the 128-ary PPM simulation accrued only 83 symbol errors over 500 Mega-symbols, & 23 bit errors over 64 Gigabits. If we run the simulations for longer, I expect the simulation results to converge to the theoretical results however running simulations of durations longer than 500 Mega-symbols is quite time consuming.

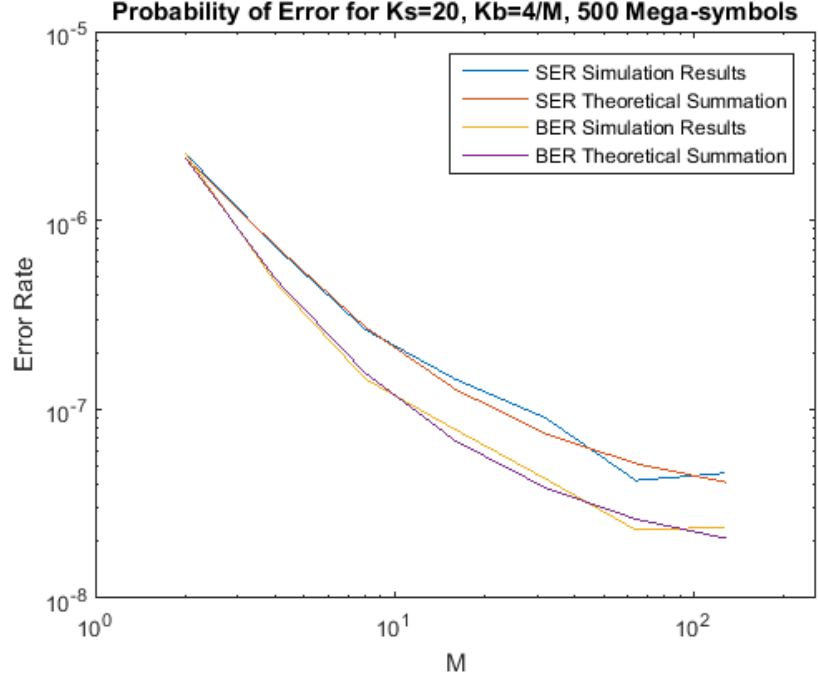

*Fig. 4 - PPM logarithmic scale theoretical versus simulation results for 500 Mega-symbols.*

Since we have now derived an expression for the BER given a symbol error for *M*-ary PPM, we can now use this expression to revisit the quantum limit:

$$BER_{Qunatum\ Limit} = \frac{M}{2(M-1)} \frac{M-1}{M} e^{-N_{ph}} = \frac{1}{2} e^{-N_{ph}}$$

Which confirms our expression from Section II.

Considering that single photon counters are not perfectly efficient devices, we can make the substitutions to our count variables:

$$K_s = N_{signal} \cdot \eta_{photon\ counter}$$

$$K_b = \frac{N_{noise} \cdot \eta_{photon\ counter} + N_{dark}}{M} \approx \frac{N_{noise} \cdot \eta_{photon\ counter}}{M}$$

Where $N_{signal}$ is the average number of signal photons arriving per symbol, $\eta_{photon\ counter}$ is the photon counting efficiency of the detector, $N_{noise}$ is the average number of noise photons arriving per symbol, & $\eta_{dark}$ is the average number of dark counts per symbol. Since the number of dark counts for high quality, modern, cooled photon counter is only in the single digits per second [10], we can effectively ignore the $N_{dark}$ term in $K_b$.

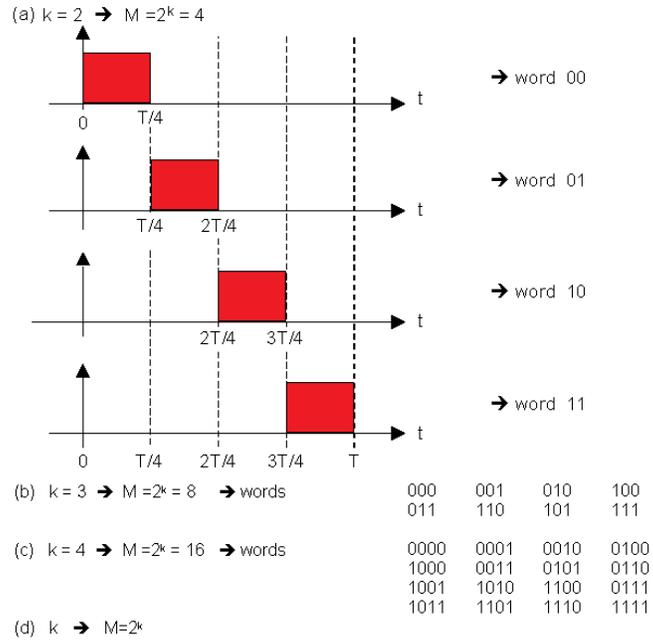

*Fig. 5 - Symbol to Bit Mappings for M-ary PPM*

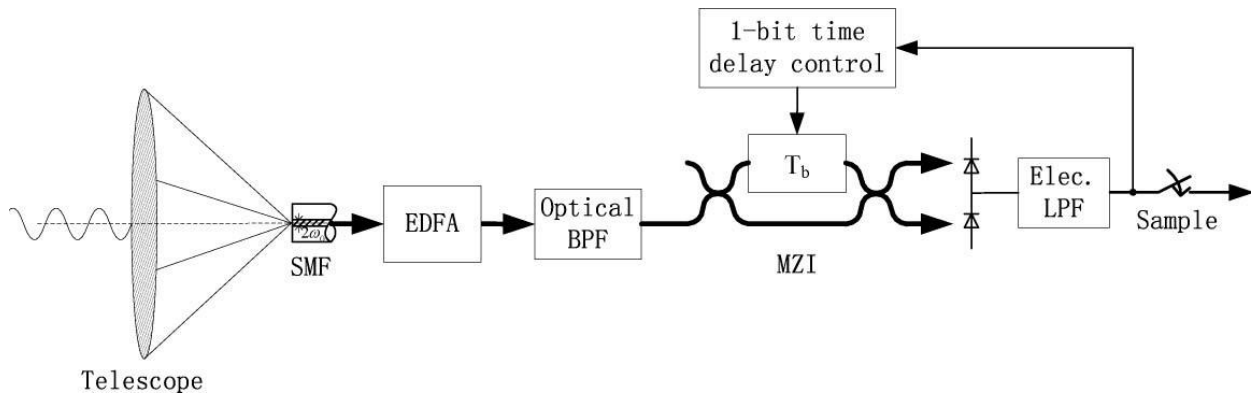

*Fig. 6 - Schematic of robust DPSK self-homodyning receiver [11].*

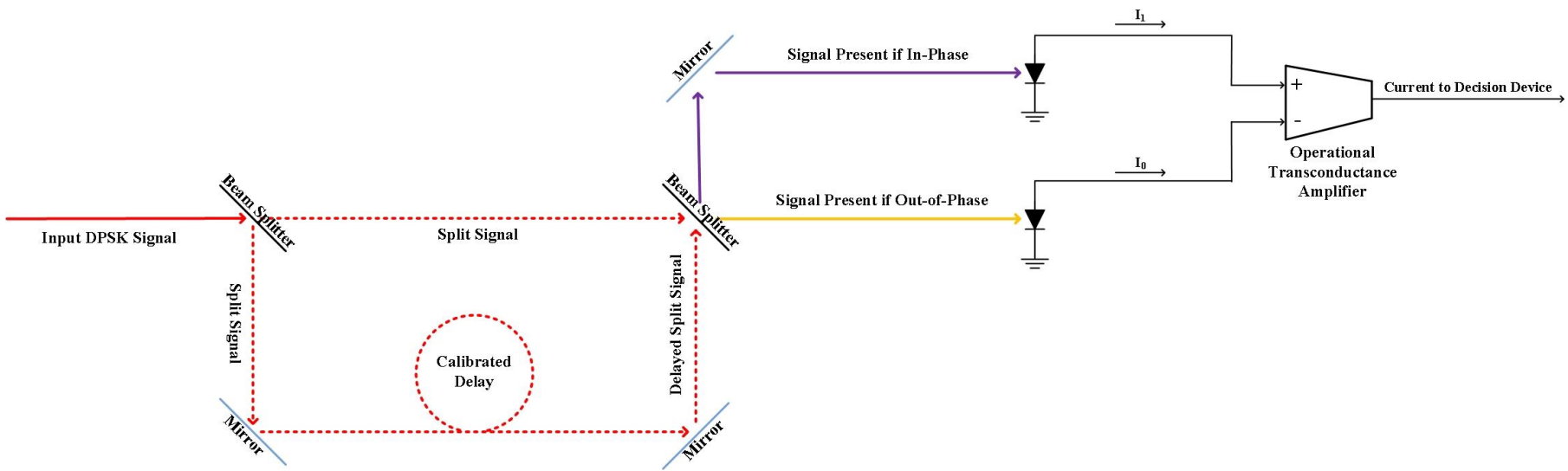

*Fig. 7 - Mach-Zehnder Interferometer with balanced detection for DPSK direct detection.*

# V. THEORETICAL PERFORMANCE OF OPTICALLY PRE-AMPLIFIED DIRECT DETECTION DPSK IN THE PRESENCE OF NOISE

NASA's LCRD project utilizes an optically pre-amplified direct detection, Mach-Zehnder interferometer based receiver for biphase DPSK supports. The basic architecture is shown in Figure 6 & Figure 7. Only 1-bit per symbol modulations are support by direct detection DPSK, though higher order modulations can be performed using heterodyne detection. The received signal is given by:

$$S_n(t) = \begin{cases} A\cos(\omega_c t + \theta) & for\ -T_b \leq t < 0 \\ A\cos(\omega_c t + \theta + n\pi) & for\ 0 \leq t < T_b \end{cases}$$

We assume that a well designed beam splitter is a material of refractive index ($n_2$) & angle of incidence ($\theta_i$) with respect to the incoming light such that the reflectance is ½, i.e. -

$$R = \left|\frac{n_1 \cos\theta_i - \sqrt{n_2^2 - n_1^2 \sin^2\theta_i}}{n_1 \cos\theta_i + \sqrt{n_2^2 - n_1^2 \sin^2\theta_i}}\right|^2 = \frac{1}{2}$$

With $R$ the reflectivity & $n_1$ the index of refraction for the media housing the beam splitters.

If the receiver is a well built Mach-Zehnder interferometer, we can analyze the input of the beam splitter using complex signal notation [12]:

$$S_{in,n}(t) = \begin{cases} \sqrt{\frac{\Phi_s}{2}} e^{j(\omega t + \theta + m\pi)} & for\ -T_b \leq t < 0 \\ \sqrt{\frac{\Phi_s}{2}} e^{j(\omega t + \theta + n\pi)} & for\ 0 \leq t < T_b \end{cases}$$

Where $\Phi_s$ is the optical power of the input signal, $\theta$ the phase, & $T_b$ the bit period. If the beam splitter is lossless, linear, & symmetric [12]:

$$\begin{bmatrix} S_1 \\ S_0 \end{bmatrix} = e^{j\theta_r} \begin{bmatrix} \sqrt{1-\mathcal{T}} & \sqrt{\mathcal{T}} e^{j\pi} \\ \sqrt{\mathcal{T}} & \sqrt{1-\mathcal{T}} \end{bmatrix} \begin{bmatrix} \sqrt{\frac{\Phi_s}{2}} e^{j(\omega t + \theta + m\pi)} \\ \sqrt{\frac{\Phi_s}{2}} e^{j(\omega t + \theta + n\pi)} \end{bmatrix}$$

$$= \sqrt{\frac{\Phi_s}{2}} e^{j(\theta_r + \theta)} \begin{bmatrix} \sqrt{1-\mathcal{T}} e^{j(\omega t + m\pi)} - \sqrt{\mathcal{T}} e^{j(\omega t + n\pi)} \\ \sqrt{\mathcal{T}} e^{j(\omega t + m\pi)} + \sqrt{1-\mathcal{T}} e^{j(\omega t + n\pi)} \end{bmatrix}$$

If we set $\mathcal{T} = \frac{1}{2}$, then $\begin{bmatrix} S_1 \\ S_0 \end{bmatrix} = \frac{\sqrt{\Phi}e^{j(\theta_r+\theta)}}{2} \begin{bmatrix} e^{j(\omega t+m\pi)} - e^{j(\omega t+n\pi)} \\ e^{j(\omega t+m\pi)} + e^{j(\omega t+n\pi)} \end{bmatrix}$ & the optical power directed at the photodiodes for 1 & 0 detection respectively is:

$$\begin{bmatrix} \Phi_1 \\ \Phi_0 \end{bmatrix} = \begin{bmatrix} S_1^* & 0 \\ 0 & S_0^* \end{bmatrix} \begin{bmatrix} S_1 \\ S_0 \end{bmatrix}$$

$$= \frac{\Phi_s}{4} \begin{bmatrix} (\cos(\omega t + m\pi) - \cos(\omega t + n\pi))^2 + (\sin(\omega t + m\pi) - \sin(\omega t + n\pi))^2 \\ (\cos(\omega t + m\pi) + \cos(\omega t + n\pi))^2 + (\sin(\omega t + m\pi) + \sin(\omega t + n\pi))^2 \end{bmatrix}$$

So, the optical power at the *1-detector* is given as:

$$\Phi_1 = \begin{cases} 0 & \text{if } m - n \text{ is even} \\ \Phi_s & \text{if } m - n \text{ is odd} \end{cases}$$

And the *0-detector* is given as:

$$\Phi_0 = \begin{cases} \Phi_s & \text{if } m - n \text{ is even} \\ 0 & \text{if } m - n \text{ is odd} \end{cases}$$

Assuming 1's & 0's are transmitted with equal probability, the average photocurrent through each diode is:

$$\begin{bmatrix} \overline{I_1} \\ \overline{I_0} \end{bmatrix} = \frac{q}{\hbar\omega} \begin{bmatrix} \eta_1 & 0 \\ 0 & \eta_0 \end{bmatrix} \begin{bmatrix} \overline{\Phi_1} \\ \overline{\Phi_0} \end{bmatrix} = \frac{q\Phi_s}{2\hbar\omega} \begin{bmatrix} \eta_1 \\ \eta_0 \end{bmatrix}$$

For photodiodes of equal quantum efficiency, $\eta_1 = \eta_0 = \eta$, &

$$\overline{I_{1,0}} = \frac{q\eta\Phi_s}{2\hbar\omega}$$

If we follow the approach of Tonguz [13], and assume amplified spontaneous emission (ASE) noise is the lead source of error for an enclosed system [13] [14]. Give ASE noise stochastic processes $n_1$ & $n_2$, with $|\sigma_{n_{1,2}}^2| = \hbar\omega \cdot n_{sp}(G-1)B$ [13]:

$$\begin{bmatrix} S_1 \\ S_0 \end{bmatrix} = \frac{e^{j\theta_r}}{2} \begin{bmatrix} 1 & -1 \\ 1 & 1 \end{bmatrix} \begin{bmatrix} \sqrt{G\Phi_s}e^{j(\omega t+\theta+m\pi)} + n_1 \\ \sqrt{G\Phi_s}e^{j(\omega t+\theta+n\pi)} + n_2 \end{bmatrix}$$

$$= \frac{e^{j\theta_r}}{2} \begin{bmatrix} \sqrt{G\Phi_s}e^{j(\omega t+\theta+m\pi)} - \sqrt{G\Phi_s}e^{j(\omega t+\theta+n\pi)} + n_1 - n_2 \\ \sqrt{G\Phi_s}e^{j(\omega t+\theta+m\pi)} + \sqrt{G\Phi_s}e^{j(\omega t+\theta+n\pi)} + n_1 + n_2 \end{bmatrix}$$

For $n - m$ an odd integer:

$$\begin{bmatrix} S_1 \\ S_0 \end{bmatrix} = \frac{e^{j(\theta_r + \omega t + \theta)}}{2} \begin{bmatrix} \sqrt{G\Phi_s} + \sqrt{G\Phi_s} + n_1' - n_2' \\ \sqrt{G\Phi_s} - \sqrt{G\Phi_s} + n_1' + n_2' \end{bmatrix}$$

$$\begin{bmatrix} \Phi_1 \\ \Phi_0 \end{bmatrix} = \begin{bmatrix} S_1^* & 0 \\ 0 & S_0^* \end{bmatrix} \begin{bmatrix} S_1 \\ S_0 \end{bmatrix} = \frac{1}{4} \begin{bmatrix} 2\sqrt{G\Phi_s}^* + n_1'^* - n_2'^* & 0 \\ 0 & n_1'^* + n_2'^* \end{bmatrix} \begin{bmatrix} 2\sqrt{G\Phi_s} + n_1' - n_2' \\ n_1' + n_2' \end{bmatrix}$$

$$= \frac{1}{4} \begin{bmatrix} 4\Phi_s + \sqrt{G\Phi_s}^*(n_1' - n_2') + \sqrt{G\Phi_s}(n_1' - n_2')^* + |n_1' - n_2'|^2 \\ |n_1' + n_2'|^2 \end{bmatrix}$$

$$= \begin{bmatrix} \Phi_s + \dfrac{\sqrt{G\Phi_s}\,\text{Real}(n_1' - n_2')}{2} + \dfrac{\text{Real}^2(n_1' - n_2')}{4} + \dfrac{\text{Imag}^2(n_1' - n_2')}{4} \\ \dfrac{\text{Real}^2(n_1' + n_2')}{4} + \dfrac{\text{Imag}^2(n_1' + n_2')}{4} \end{bmatrix}$$

$$= \begin{bmatrix} \left(\sqrt{G\Phi_s} + \dfrac{\text{Real}(n_1' - n_2')}{2}\right)^2 + \dfrac{\text{Imag}^2(n_1' - n_2')}{4} \\ \dfrac{\text{Real}^2(n_1' + n_2')}{4} + \dfrac{\text{Imag}^2(n_1' + n_2')}{4} \end{bmatrix}$$

The output current will be:

$$I_{out} = \Re(\Phi_1 - \Phi_0)$$

One can create a decision voltage for each bit of:

$$V_{decision} = \frac{Q_{out}}{C} = \frac{1}{C} \int_{m \cdot T_b + t_0}^{(m+1) \cdot T_b + t_0} I_{out}\, dt$$

$$= \frac{\Re}{C} \left\{ \int_{m \cdot T_b + t_0}^{(m+1) \cdot T_b + t_0} \left(\sqrt{G\Phi_s} + \frac{\text{Real}(n_1' - n_2')}{2}\right)^2 + \frac{\text{Imag}^2(n_1' - n_2')}{4}\, dt \right.$$

$$\left. - \left( \int_{m \cdot T_b + t_0}^{(m+1) \cdot T_b + t_0} \frac{\text{Real}^2(n_1' + n_2')}{4} + \frac{\text{Imag}^2(n_1' + n_2')}{4}\, dt \right) \right\}$$

We have two integrals of stochastic processes the 1st non-central having a non-central chi-squared distributions & the 2nd a central chi-squared distributions. The result is a difference of chi-squared distributions defined by:

$$V_{decision} = \frac{\Re}{C}\left\{\left(\sqrt{G\Phi_s T_b} + w_1\right)^2 + w_2^2 - (w_3^2 + w_4^2)\right\}$$

Where $w_1$ through $w_4$ are stochastic processes with Gaussian PDFs & the variance of $\frac{\hbar\omega \cdot n_{sp}(G-1)B \cdot T_b}{4}$, where $n_{sp}$ is the spontaneous emission parameter of the amplifying medium, $G$ the gain of the amplifier, & $B$ the bandwidth. For $G \gg 1$ & $B = R_b$, $\sigma^2_{w_{1,2,3,4}} = \frac{\hbar\omega n_{sp} G}{4}$. The probability of error is:

$$P_e = Prob\left(\left(\sqrt{G\Phi_s T_b} + w_1\right)^2 + w_2^2 < w_3^2 + w_4^2\right)$$

At this point, we borrow a trick from Carlson [15], & write the equivalent probability:

$$P_e = Prob\left(\sqrt{\left(\sqrt{G\Phi_s T_b} + w_1\right)^2 + w_2^2} < \sqrt{w_3^2 + w_4^2}\right)$$

Or the probability that the vectors $\begin{pmatrix}\sqrt{G\Phi_s T_b} + w_1 \\ w_2\end{pmatrix}$ & $\begin{pmatrix}w_3 \\ w_4\end{pmatrix}$, $\left|\begin{matrix}\sqrt{G\Phi_s T_b} + w_1 \\ w_2\end{matrix}\right| < \left|\begin{matrix}w_3 \\ w_4\end{matrix}\right|$. In other words, the probability of error is the probability that the line from $\begin{pmatrix}0\\0\end{pmatrix}$ to $\begin{pmatrix}\sqrt{G\Phi_s T_b} + w_1 \\ w_2\end{pmatrix}$ is shorter than the line to $\begin{pmatrix}0\\0\end{pmatrix}$ to $\begin{pmatrix}w_3 \\ w_4\end{pmatrix}$. $\left|\begin{matrix}\sqrt{G\Phi_s T_b} + w_1 \\ w_2\end{matrix}\right| = \sqrt{\left(\sqrt{G\Phi_s T_b} + w_1\right)^2 + w_2^2}$ has a Rician PDF described by $\frac{x}{\sigma^2} e^{\frac{-(x^2-v^2)}{2\sigma^2}} I_0\left(\frac{xv}{\sigma^2}\right)$ where $\sigma^2 = \frac{\hbar\omega n_{sp}(G-1)}{2}$, $v = \sqrt{G\Phi_s T_b}$, & $I_0$ is a modified Bessel function of the 1st kind. $\left|\begin{matrix}w_3 \\ w_4\end{matrix}\right| = \sqrt{w_3^2 + w_4^2}$ has a Rayleigh PDF described by $\frac{x}{\sigma^2} e^{\frac{-(x^2-v^2)}{2\sigma^2}} I_0\left(\frac{xv}{\sigma^2}\right)$. The probability of error is therefore given by:

$$P_e = Prob\left(\sqrt{\left(\sqrt{G\Phi_s T_b} + w_1\right)^2 + w_2^2} < \sqrt{w_3^2 + w_4^2}\right)$$

$$= \int_0^\infty \frac{x_1}{\sigma^2} e^{\frac{-(x_1^2+v^2)}{2\sigma^2}} I_0\left(\frac{x_1 v}{\sigma^2}\right) \left(\int_{x_1}^\infty \frac{x_2}{\sigma^2} e^{\frac{-x_2^2}{2\sigma^2}} dx_2\right) dx_1$$

$$= \int_0^\infty \frac{x_1}{\sigma^2} e^{\frac{-(x_1^2+v^2)}{2\sigma^2}} I_0\left(\frac{x_1 v}{\sigma^2}\right) \left(\int_{\frac{x_1^2}{2\sigma^2}}^\infty e^{-x_2'} dx_2'\right) dx_1$$

$$= \int_0^\infty \frac{x_1}{\sigma^2} e^{\frac{-(x_1^2+v^2)}{2\sigma^2}} I_0\left(\frac{x_1 v}{\sigma^2}\right) e^{\frac{-x_1^2}{2\sigma^2}} dx_1 = \int_0^\infty \frac{x_1}{\sigma^2} e^{\frac{-(2 \cdot x_1^2+v^2)}{2\sigma^2}} I_0\left(\frac{x_1 v}{\sigma^2}\right) dx_1$$

Setting $x_1' = x_1\sqrt{2}$ & $v' = \frac{v}{\sqrt{2}}$,

$$P_e = \int_0^\infty \frac{x_1}{\sigma^2} e^{\frac{-(2\cdot x_1^2 + v^2)}{2\sigma^2}} I_0\left(\frac{x_1 v}{\sigma^2}\right) dx_1 = \int_0^\infty \frac{x_1'}{\sigma^2 \sqrt{2}} e^{\frac{-(x_1'^2 + 2v'^2)}{2\sigma^2}} I_0\left(\frac{x_1' v'}{\sigma^2}\right) dx_1$$

$$= \frac{1}{2} e^{\frac{-v'^2}{2\sigma^2}} \int_0^\infty \frac{x'}{\sigma^2} e^{\frac{-(x_1'^2 + v'^2)}{2\sigma^2}} I_0\left(\frac{x_1' v'}{\sigma^2}\right) dx'$$

$\frac{x'}{\sigma^2} e^{\frac{-(x_1'^2 + v'^2)}{2\sigma^2}} I_0\left(\frac{x_1' v'}{\sigma^2}\right)$ is just a Rician PDF which integrates to one so,

$$P_e = \frac{1}{2} e^{\frac{-v'^2}{2\sigma^2}} = \frac{1}{2} e^{\frac{-v^2}{4\sigma^2}} = \frac{1}{2} e^{\frac{-G\Phi_s T_b}{\hbar\omega(G-1)n_{sp}}}$$

Since $\Phi_s = \hbar\omega N_{ph} R_b$, the probability of error is simply:

$$P_e = \frac{1}{2} e^{\frac{-G\cdot N_{ph}}{(G-1)\cdot n_{sp}}} \approx \frac{1}{2} e^{\frac{-N_{ph}}{n_{sp}}} \; for \; high \; gain$$

Where $n_{sp}$ is the stimulated emission coefficient [14], given by:

$$n_{sp} = \frac{N_2}{N_2 - N_1}$$

$N_1, N_2$ are the number of lasing atoms in the excited and ground states, respectively, in the population inverted laser amplifier.

It is noteworthy to mention that shot noise for the optically pre-amplified condition has been ignored from in this analysis, because it is expected that for large gain $(G-1)\cdot n_{sp} \gg \sqrt{N_{ph}}$, where the left hand side is the would be shot noise term. A more thorough handling of shot noise is considered in the heterodyne DPSK BER derivation given in Appendix A. In this analysis we have neglected external noise, in part to verify the result of our analysis matches that given by Tonguz [13]. In the presence of interfering in-band light with expected in-polarization photon count per symbol of $N_b$, our probability of error is easily modified:

$$P_e = \frac{1}{2} e^{\frac{-N_{ph}}{\frac{G-1}{G} n_{sp} + N_b}}$$

Clearly, in the absence of noise this & high gain, we can write a quantum limit for DPSK of $\frac{1}{2} e^{-N_{ph}}$, which matches what was derived in the PPM section.

## VI. COMPARISON BETWEEN PPM & DPSK IN A REALISTIC SCENERIO

From the symbolic math presented thus far it is difficult to glean any insight as to what modulation type is superior, as aside from the quantum limited cases which are equivalent, the bit error rate formula for PPM is presented as a large summation & the bit error rate formula for DPSK is given as an exponential. The 1$^{st}$ order of business is to generate reasonable values for the number of unwanted noise photons entering the optical communications ground terminal's receiving telescope system. The LCRD mission operates at optical C-band [2], where the band center is 1547.5 nm. Using this data point, investigate the typical luminescence of this infrared (IR) wavelength in the sky.

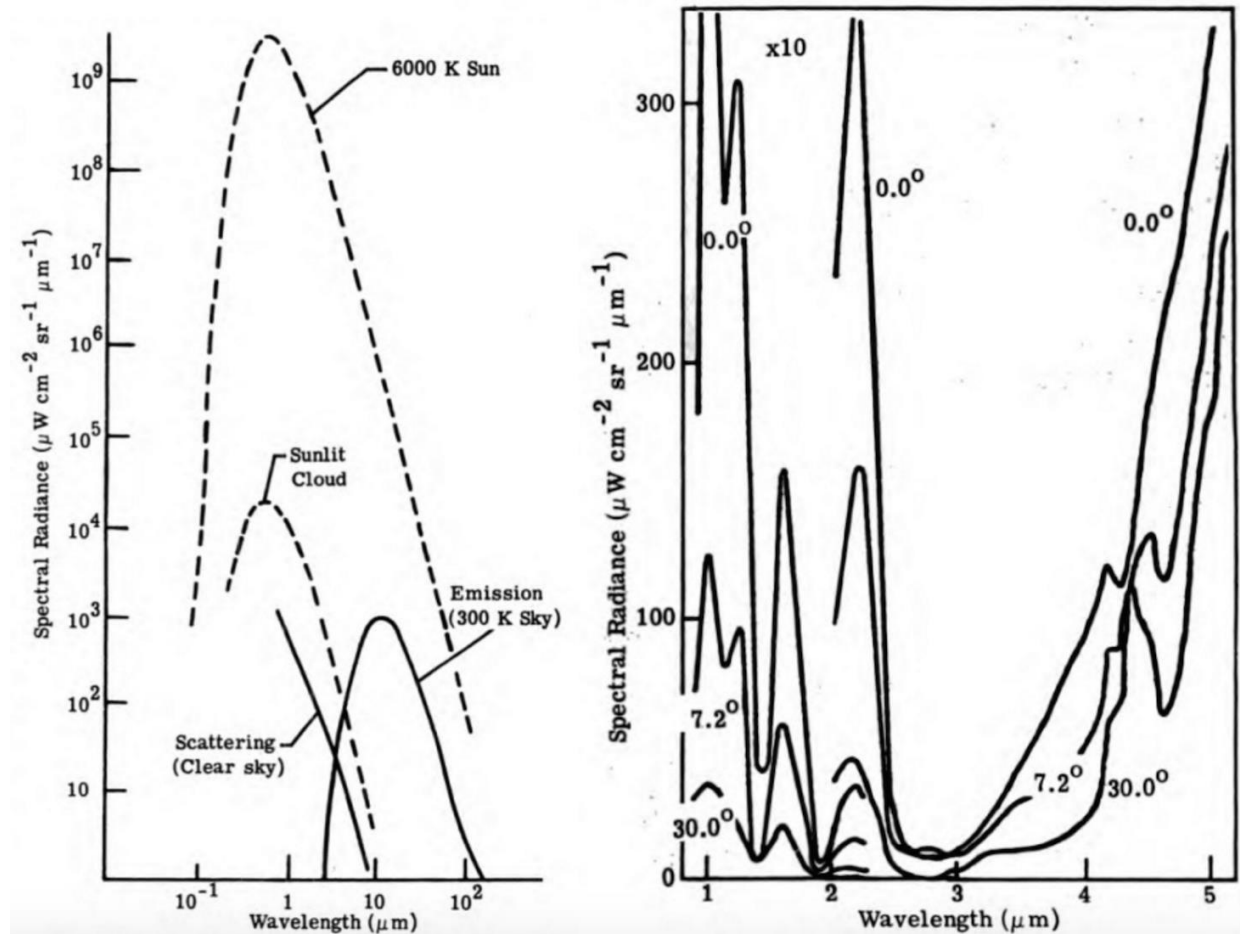

*Fig. 8* - **Contributions of scattering and atmospheric emission to background radiation (left), and the spectral radiance of a clear daytime sky, both taken from** *The Infrared Handbook* **[16].**

A good estimate of the noise power in $\frac{\mu W}{cm^2 \, sr \, \mu m}$ seems to be 10 or lower, at C-band. A telescope aperture diameter of roughly 40 cm is typical in the optical link budgets the author has seen, and optical filters with bandwidths of 0.5 nm are commercially available [17]. Finally, we can derive a solid angle for our noise computation using the formula:

$$\Omega = 2\pi(1 - \cos(\theta)) \, sr$$

Where $\theta$ is the angular diameter of the telescope view. Substituting $0.5° \frac{\pi}{180°}$ for $\theta$ yields a solid angle of $59.8 \, \mu sr$. For a spectral radiance of $10 \, \frac{\mu W}{cm^2 \, sr \, \mu m}$, we calculate a noise power at the aperture of $10 \cdot \pi \cdot 20^2 \, 59.8 \cdot 10^{-6} \cdot .5 \cdot 10^{-3} = 3.76 \cdot 10^{-4} \mu W = 37.6 \, pW$. Dividing by *hf* at our frequency of interest yields 2.93 billion photons/second.

At this stage, the data rate of the system becomes of key importance. 311 Mbps is a common data rate for both DPSK & PPM in the LCRD requirements, so for a fair comparison of the 2 modulation schemes we choose this value. Clearly, for DPSK number of noise photons per symbol is simply the number of noise photons entering the system per second times the bit period, which is the reciprocal of the data rate. This yields a background count of 9.41 photons per symbol, however assuming we have a polarization filter & are receiving only 1 polarization we can divide by factor of 2, yielding 4.71 photons per symbol in polarization. Note that this is the highest noise level that we will analyze, so the background count should seem a bit on the high side. For PPM, the number of background counts per slot will be determined by the symbol rate of $\frac{311 \, Mbps}{log_2(M)}$ divided by the number of slots per symbol, $M$.

Applying a photon counting efficiency of 70% for our high speed photon counter, & for our DPSK pre-amplifier we assign a gain of 40 dB & $n_{sp} = 1.05$. With this we can calculate a bit error rate curve for our test-case in MATLAB.

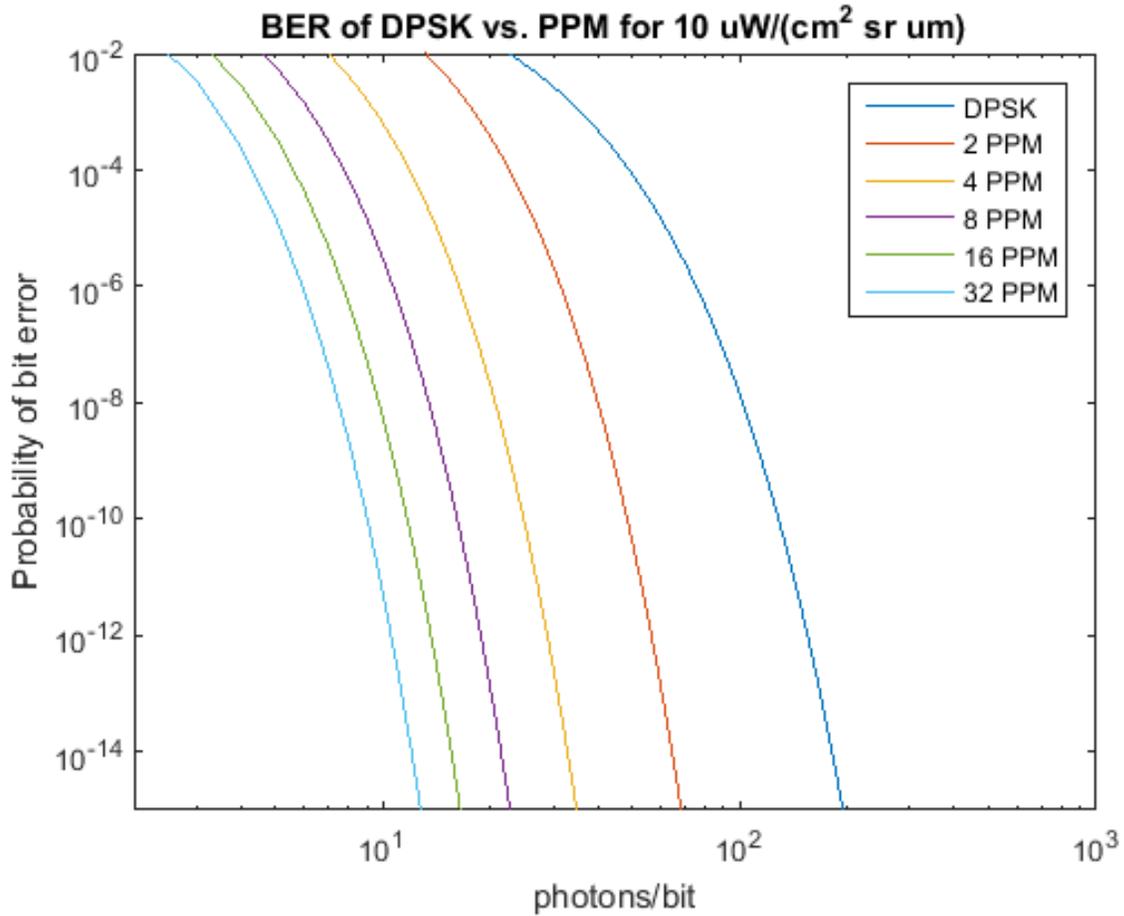

*Fig. 9* - **BER Curve for 311 Mbps, 10 $\frac{\mu W}{cm^2\,sr\,\mu m}$ case.**

| M | Symbol Rate (Msps) | Background photons per symbol in polarization | Background photons per slot |
|---|---|---|---|
| DPSK | 311 | 4.706817423 | 4.706817423 |
| 2 | 311 | 4.706817423 | 2.353408712 |
| 4 | 155.5 | 9.413634846 | 2.353408712 |
| 8 | 103.6666667 | 14.12045227 | 1.765056534 |
| 16 | 77.75 | 18.82726969 | 1.176704356 |
| 32 | 62.2 | 23.53408712 | 0.735440222 |

*Table 1* - **Background photon breakdown for different values of *M* for the PPM 311 Mbps, 10 $\frac{\mu W}{cm^2\,sr\,\mu m}$ case before applying efficiency corrections.**

For the high-noise case, we see that DPSK is the clear loser, with PPM performing better and better which each higher value of *M*. We now reduce our noise level by 3 decibels, to another reasonable point on the *Infrared Handbook* curve.

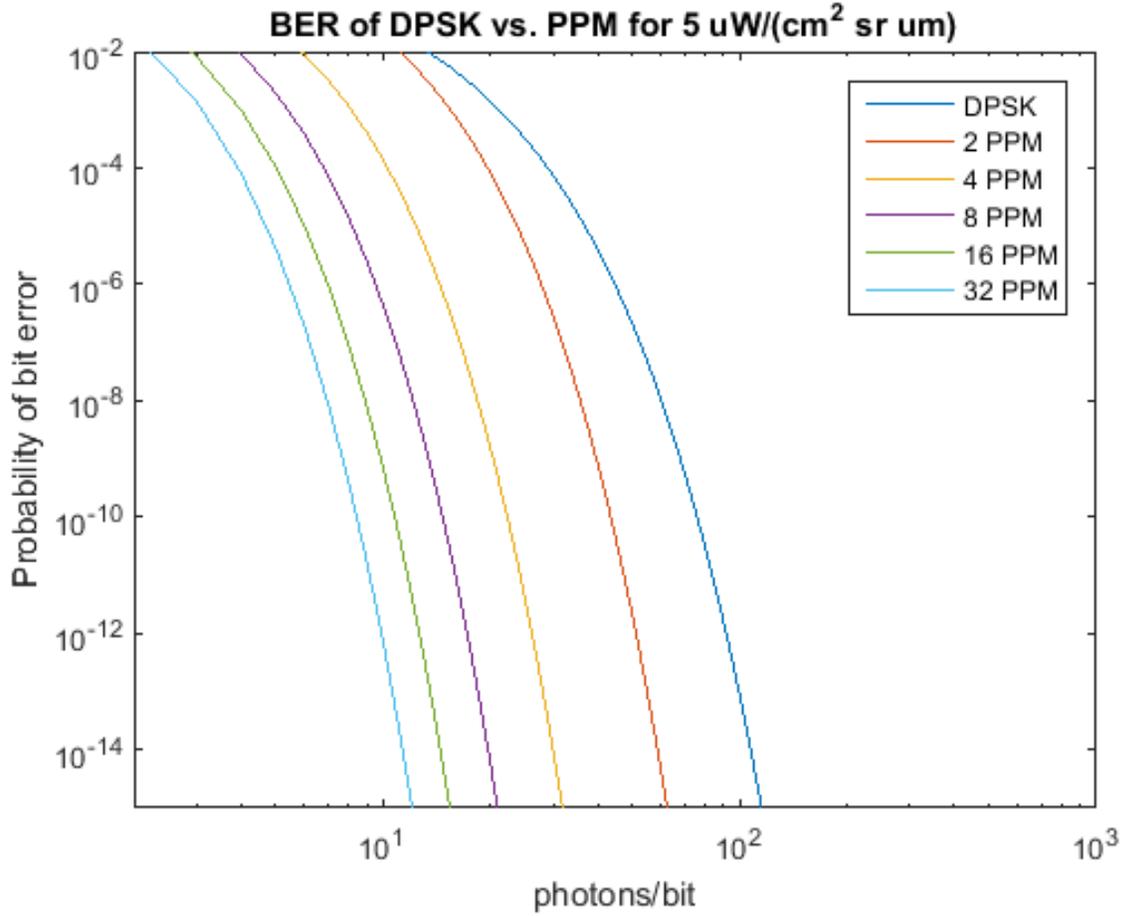

*Fig. 10* - **BER Curve for 311 Mbps, 5 $\frac{\mu W}{cm^2\ sr\ \mu m}$ case.**

| M | Symbol Rate (Msps) | Background photons per symbol in polarization | Background photons per slot |
|---|---|---|---|
| DPSK | 311 | 2.353408712 | 2.353408712 |
| 2 | 311 | 2.353408712 | 1.176704356 |
| 4 | 155.5 | 4.706817423 | 1.176704356 |
| 8 | 103.6666667 | 7.060226135 | 0.882528267 |
| 16 | 77.75 | 9.413634846 | 0.588352178 |
| 32 | 62.2000002 | 11.76704352 | 0.36772011 |

*Table 2* - **Background photon breakdown for different values of *M* for the PPM 311 Mbps, 5 $\frac{\mu W}{cm^2\ sr\ \mu m}$ case before applying efficiency corrections.**

We now see that all of the curves have pulled in, as expected, with DPSK seeing the most improvement. We generate some additional plots for the reader to consider, below, with lower values for spectral radiance, since these systems will also operate at night.

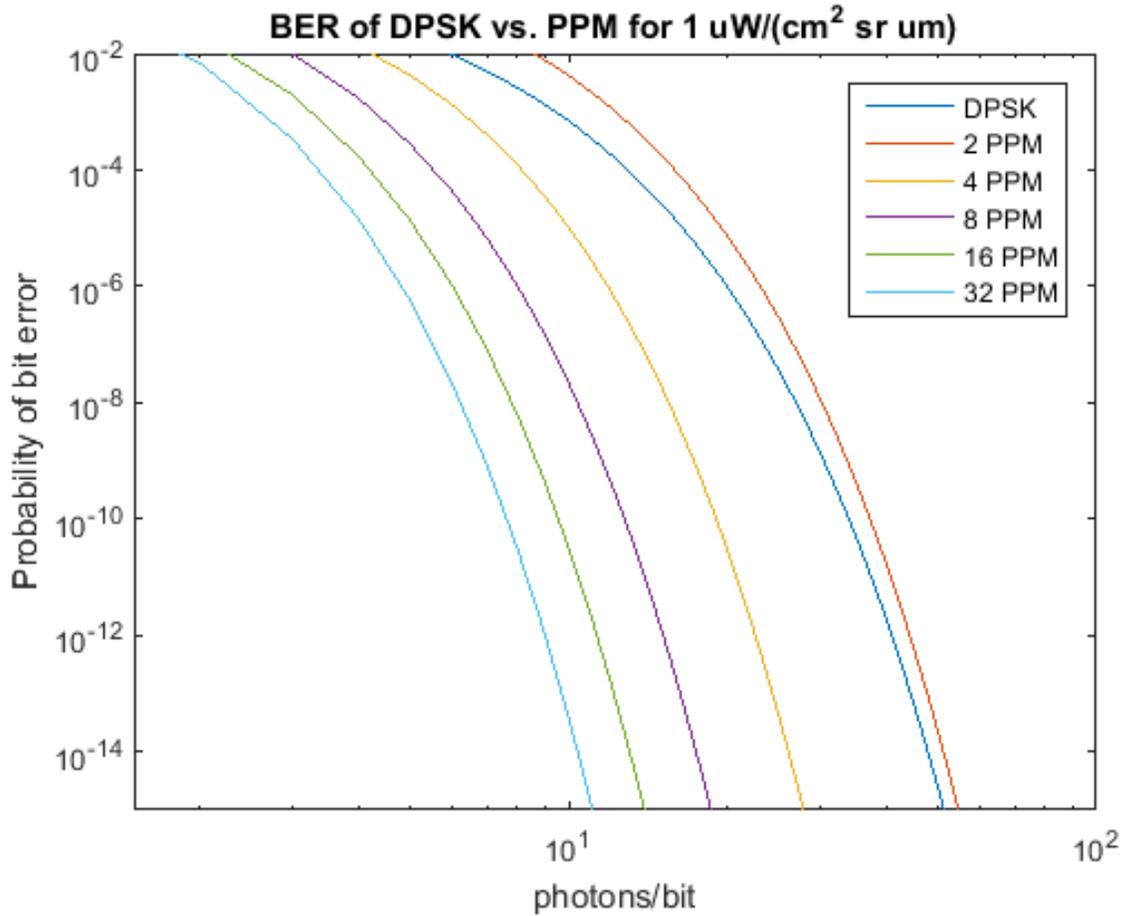

*Fig. 11* - BER Curve for 311 Mbps, $1 \frac{\mu W}{cm^2\ sr\ \mu m}$ case.

| M | Symbol Rate (Msps) | Background photons per symbol in polarization | Background photons per slot |
|---|---|---|---|
| DPSK | 311 | 0.470681742 | 0.470681742 |
| 2 | 311 | 0.470681742 | 0.235340871 |
| 4 | 155.5 | 0.941363485 | 0.235340871 |
| 8 | 103.6666667 | 1.412045227 | 0.176505653 |
| 16 | 77.75 | 1.882726969 | 0.117670436 |
| 32 | 62.2000002 | 2.353408704 | 0.073544022 |

*Table 3* - Background photon breakdown for different values of *M* for the PPM 311 Mbps, $1 \frac{\mu W}{cm^2\ sr\ \mu m}$ case before applying efficiency corrections.

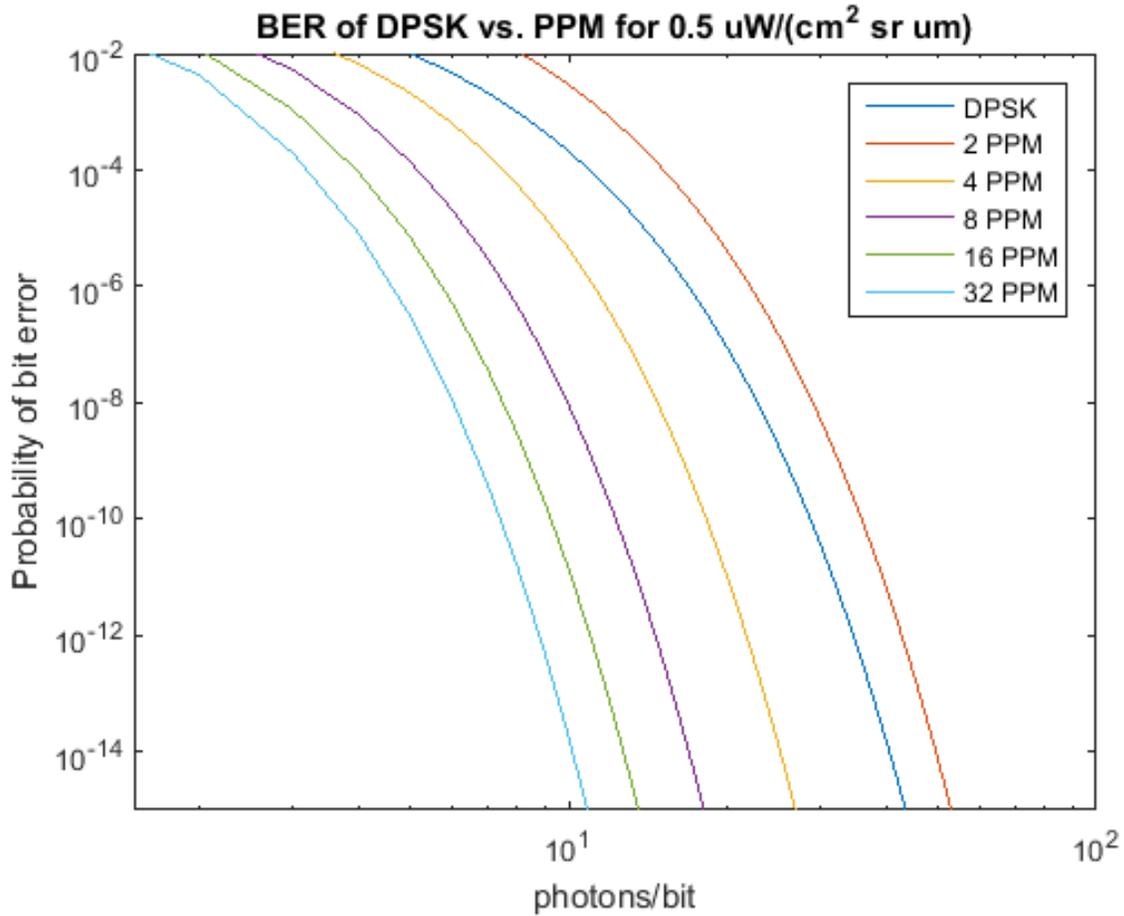

*Fig. 12* - BER Curve for 311 Mbps, $0.5 \frac{\mu W}{cm^2 \, sr \, \mu m}$ case.

| M | Symbol Rate (Msps) | Background photons per symbol in polarization | Background photons per slot |
|---|---|---|---|
| DPSK | 311 | 0.235340871 | 0.235340871 |
| 2 | 311 | 0.235340871 | 0.117670436 |
| 4 | 155.5 | 0.470681742 | 0.117670436 |
| 8 | 103.6666667 | 0.706022613 | 0.088252827 |
| 16 | 77.75 | 0.941363485 | 0.058835218 |
| 32 | 62.2000002 | 1.176704352 | 0.036772011 |

*Table 4* - Background photon breakdown for different values of *M* for the PPM 311 Mbps, $0.5 \frac{\mu W}{cm^2 \, sr \, \mu m}$ case before applying efficiency corrections.

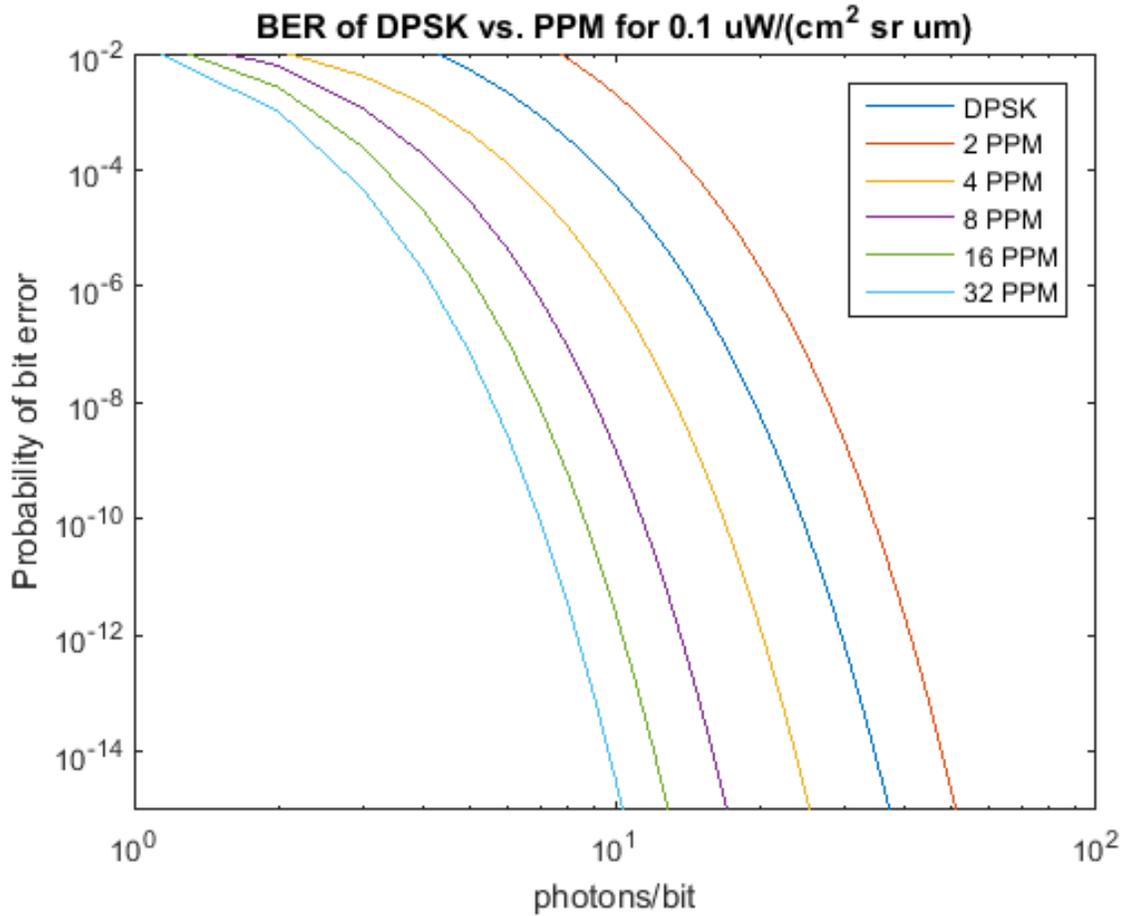

*Fig. 13* - BER Curve for 311 Mbps, $0.1 \frac{\mu W}{cm^2 \, sr \, \mu m}$ case.

| M | Symbol Rate (Msps) | Background photons per symbol in polarization | Background photons per slot |
|---|---|---|---|
| DPSK | 311 | 0.047068174 | 0.047068174 |
| 2 | 311 | 0.047068174 | 0.023534087 |
| 4 | 155.5 | 0.094136348 | 0.023534087 |
| 8 | 103.6666667 | 0.141204523 | 0.017650565 |
| 16 | 77.75 | 0.188272697 | 0.011767044 |
| 32 | 62.2000002 | 0.23534087 | 0.007354402 |

*Table 5* - Background photon breakdown for different values of *M* for the PPM 311 Mbps, $0.1 \frac{\mu W}{cm^2 \, sr \, \mu m}$ case before applying efficiency corrections.

## VII. DISCUSSION

The higher order PPM (*M* of 4 and about) modes are superior to DPSK in every scenario that we have analyzed in the previous section. The LCRD project has elected to use 2 *M* values in their project requirements: 4-ary PPM for forward (transmit) services and 16-ary PPM for return (receive) services. Though the forward service noise parameters should be somewhat different than in our ground telescope centered analysis, we can still make the qualitative judgment that performance as measured by BER at a given number of average received photons per bit will be better for each of the PPM cases than for DPSK, with performance improving for each higher value of *M*.

In our analysis, however, we have only analyzed the performance based off the average number of photons received per bit, and have not considered laser power usages when the PPM receiver is not transmitting. A Q-switched laser used to generate the short pulses transmitting the information for PPM is not using power exclusively when transmitting. As an example, if we assume the Q-switching is only 50% efficient, which seems reasonable given that the laser must be able to generate 2 consecutive pulses given an $M^{th}$ position symbol followed by a first position symbol, then the BER for the DPSK case actually overtakes the 4-ary PPM case at $1\ \frac{\mu W}{cm^2\ sr\ \mu m}$ and lower spectral radiance levels for reasonable bit error rates.

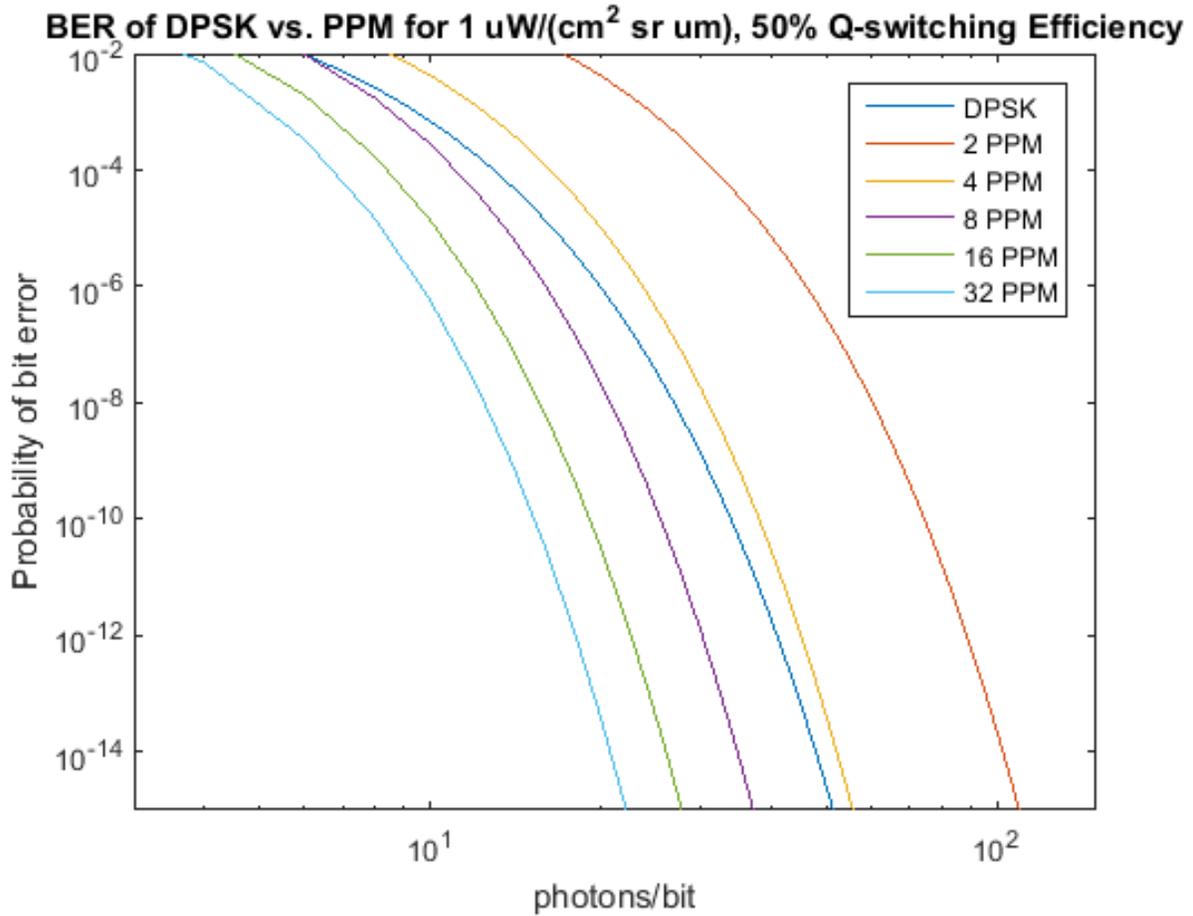

*Fig. 14* - **BER Curve for 311 Mbps, 1 $\frac{\mu W}{cm^2\ sr\ \mu m}$ case when a 50% Q-switching power efficiency is applied to the number of photons for the PPM cases.**

It is clearly demonstrated by examining the figures in Section VI that DPSK is more sensitive to external noise. This is in part due to the fact that the only efficiency consideration taking for DPSK was the relatively modest spontaneous emission factor of 1.05 was applied, whereas a photon counting efficiency of 70% was applied to the PPM calculations. Low photon counting efficiency for PPM hurts performance overall, however also reduces the probability that noise photons are counted. In general, higher photon counting efficiency improves the PPM bit error rate. To demonstrate this, we include an idealized curve as Figure 15 where both the spontaneous emission parameter & the photon counting efficiency are set to 1 (100%), and also the gain of the

DPSK pre-amplifier is assume to be infinite ($\frac{G-1}{G} = 1$). For the 1 $\frac{\mu W}{cm^2\ sr\ \mu m}$ moderate case, 2-ary PPM has gone from being marginally worse than DPSK to being more efficient.

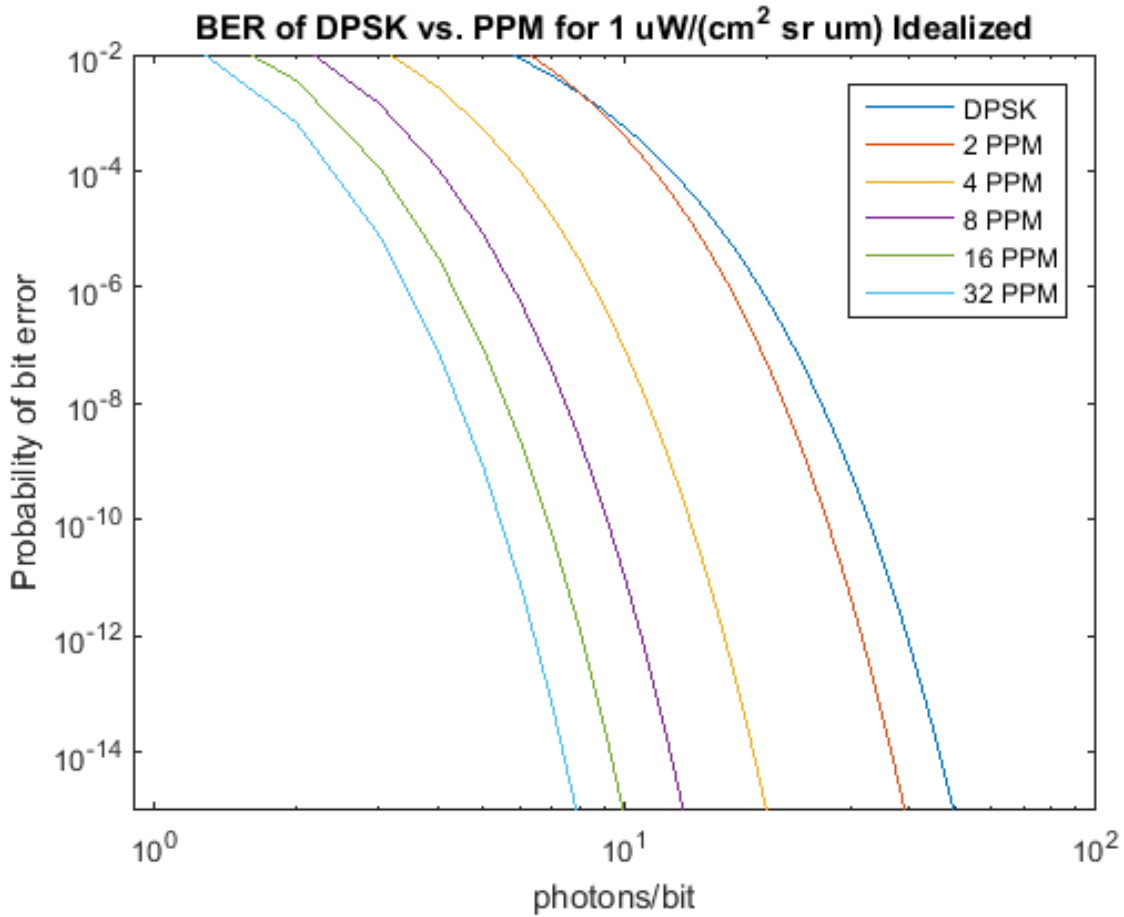

*Fig. 15* - **BER Curve for 311 Mbps, 1 $\frac{\mu W}{cm^2\ sr\ \mu m}$ case when with idealized photon counting efficiency, DPSK pre-amplifier spontaneous emission parameter, & DPSK pre-amplifier gain.**

So, if higher order PPM performance is so much better than DPSK than why is NASA investing in DPSK technologies? The answer is that, in the short term, DPSK ultimately promises to provide higher data rate capabilities than achievable with PPM, with over 10 Gbps on the near term horizon [18]. Ultimately, photon counters cannot detect an event and reset in an arbitrary amount of time & also only provide finite timing resolution. For DPSK, the output of the hardware receiver can be sampled by a very fast analog-to-digital converter, which does not have the requirement to reset the photon counter for each received photon. The motivation behind DPSK technology development is not efficiency, but higher data rates.

The analysis thus presented has been instructed, however it is worthwhile to discuss considerations for performance that we have ignored. We have not discussed the effect of error correction coding in this analysis, which will vastly improve the BER performance for every case, nor have we discussed acquisition time, which is a key performance parameter in the satellite communications arena. To acquire and successfully extract data from a signal which utilizes a high-quality block coding scheme, such as low-density parity-check codes (LDPC), one must first search for & lock to a frame synchronization marker using a search sequence. Additionally, issues such as phase noise & Doppler shift are known to affect DPSK BER performance [11], but we save discussion of the contributions of these affects for a future work. Finally, in the case of DPSK we have trusted the outputs of the relevant literature [13] [14] in assuming ASE noise is the limiting factor in BER of non-free space DPSK systems, and a more thorough analysis would include thermal noise, flicker noise, phase noise, & possibly additional contributions.

## VIII. CONCLUSION

Based off this analysis, higher order PPM has superior BER performance as compared to DPSK for reasonable BER operating points. Other factors, however, motivate the development of DPSK for satellite communications, such as peak data rate. These observations are in agreement with those stated by NASA [2] [18], so we can assume this analysis is generally correct. BER performance of optical technologies can be improved further by the development of higher order PPM technologies using better speed & resolution photon counting systems, as well as photon counters with higher efficiency, as discussed in the section VII.

# APPENDIX A:
# BER PERFORMANCE HETERODYNE DETECTION DPSK

It is somewhat novel that analysis of heterodyne DPSK receiver systems yields a similar result to that shown in the homodyne, direct detection case so we include this section to illustrate what has been, in the past, a point of confusion [13]. After all, for direct detection one is adding 2 signals together using constructive and destructive interference & with heterodyne systems one is multiplying the signals together. So far, through the use of Poisson statistics & the novel characteristics of optical pre-amplifiers, we have avoided an in-depth analysis of shot noise. In order analyze the theoretical limits of the heterodyne DPSK optical receiver, we must take an in-depth look at this noise source as it will become a limiting factor.

If we assume that the receiving system is well engineered, is sufficiently optically pre-amplified, & the electro-optic components do not significantly contribute to bit errors in the detector. In this case the system's noise will be solely due to power fluctuations on the incoming light (photon shot noise). Considering the incoming light as a stream of photons, each photon does not travel through space as either a time dependent Dirac delta function nor a rect function such that the photon energy density was constant over some interval, but instead will be approximately Gaussian around some time, $t$, we can write the photon's energy density function as a function of time (the power):

$$\Phi(t) = \frac{\hbar\omega}{\sigma\sqrt{2\pi}} e^{-\frac{t^2}{2\sigma^2}}$$

Where $\Phi$ is the optical signal power & $\sigma$ the standard deviation. Clearly, $\int_{-\infty}^{\infty} \Phi(t) dt = \hbar\omega$, the energy per photon. The Fourier transform of the optical power is:

$$\Phi(\omega) = \int_{-\infty}^{\infty} \Phi(t) e^{-j\omega t} dt = \frac{\hbar\omega}{\sigma\sqrt{2\pi}} \int_{-\infty}^{\infty} e^{-\frac{t^2}{2\sigma^2} - j\omega t} dt = \frac{\hbar\omega}{\sigma\sqrt{2\pi}} e^{-\frac{\sigma^2\omega^2}{2}} \int_{-\infty}^{\infty} e^{-\left(\frac{t}{\sigma\sqrt{2}} + j\frac{\sigma\omega}{\sqrt{2}}\right)^2} dt$$

$$= \frac{\hbar\omega}{\sqrt{\pi}} e^{-\frac{\sigma^2\omega^2}{2}} \int_{-\infty - j\frac{\sigma\omega}{\sqrt{2}}}^{\infty - j\frac{\sigma\omega}{\sqrt{2}}} e^{-t'^2} dt'$$

Where $t' = \frac{t}{\sigma\sqrt{2}} + j\frac{\sigma\omega}{\sqrt{2}}$. Since:

$$\int_{-\infty-j\frac{\sigma\omega}{\sqrt{2}}}^{\infty-j\frac{\sigma\omega}{\sqrt{2}}} e^{-t'^2} dt' = \sqrt{\int_{-\infty-j\frac{\sigma\omega}{\sqrt{2}}}^{\infty-j\frac{\sigma\omega}{\sqrt{2}}} \int_{-\infty-j\frac{\sigma\omega}{\sqrt{2}}}^{\infty-j\frac{\sigma\omega}{\sqrt{2}}} e^{-(x^2+y^2)} dx\, dy} = \sqrt{\int_0^{2\pi} \int_0^\infty e^{-r^2} r\, dr\, d\theta}$$

$$= \sqrt{2\pi \left(\frac{1}{2}\int_0^\infty e^{-r^2} 2r\, dr\right)} = \sqrt{\pi \cdot -(e^{-\infty^2} - e^{-0^2})} = \sqrt{\pi}$$

We have:

$$\Phi(\omega) = \hbar\omega e^{-\frac{\sigma^2\omega^2}{2}}$$

Which is another Gaussian. For $\frac{\sigma^2\omega^2}{2} \ll 0.001$ the Fourier transform is is effectively flat [19], & $\hbar\omega$ is the 2-sided power spectral density associated with 1 photon. As such $2\hbar\omega$ is the 1-sided power spectral density. The variance of the power associated with 1 photon is therefore $\hbar^2\omega^2$ akin to how the variance of an RF signal (as opposed to the power) is half the one-sided noise power spectral density, or $\frac{n_0}{2}$ [9]. The typical interpretation of this is that [19] randomly occurring pulses of light contribute broadband white noise. To see the effect of this on a photo-diode driving a load resistant, R, we note that $i(t) = \Re\Phi(t)$, with $\Re = \frac{\eta q}{\hbar\omega}$ the responsivity of the diode. Note $q$ is the elementary charge, & $\eta$ the quantum efficiency of the photodiode. The average power through the resistor being driven over some time T is:

$$P = \frac{1}{T}\int_{-\frac{T}{2}}^{\frac{T}{2}} R \cdot i^2(t) dt = \frac{R\Re^2}{T}\int_{-\frac{T}{2}}^{\frac{T}{2}} \Phi^2(t) dt = \frac{R\Re^2}{T}\int_{-\frac{T}{2}}^{\frac{T}{2}} \Phi(t) \left[\frac{1}{2\pi}\int_{-\infty}^{\infty} \Phi(\omega) e^{j\omega t} d\omega\right] dt$$

Where we have used an inverse Fourier transform of the optical power. If we assume that the photon's arrival is well localized around time $t = 0$, & negligible power arrives outside of $\pm\frac{T}{2}$, we can write:

$$P = \frac{R\Re^2}{2\pi T}\int_{-\infty}^{\infty}\int_{-\infty}^{\infty} \Phi(t)\Phi(\omega) e^{j\omega t} d\omega dt = \frac{R\Re^2}{2\pi T}\int_{-\infty}^{\infty} \Phi(\omega) \left[\int_{-\infty}^{\infty} \Phi(t) e^{j\omega t} dt\right] d\omega$$

Since $\Phi(t)$ is real,

$$\int_{-\infty}^{\infty} \Phi(t)e^{j\omega t}dt = \left[\int_{-\infty}^{\infty} \Phi(t)[e^{j\omega t}]^*dt\right]^* = \left[\int_{-\infty}^{\infty} \Phi(t)e^{-j\omega t}dt\right]^* = \Phi^*(\omega)$$

$$\therefore P = \frac{R\mathfrak{R}^2}{2\pi T}\int_{-\infty}^{\infty} \Phi(\omega)\,\Phi^*(\omega)d\omega = \frac{R\mathfrak{R}^2}{2\pi T}\int_{-\infty}^{\infty} |\Phi(\omega)|^2\,d\omega = \frac{R\mathfrak{R}^2}{\pi T}\int_{0}^{\infty} |\Phi(\omega)|^2\,d\omega$$

The power spectral density is then:

$$P(\omega) = \frac{R|\Phi(\omega)|^2 \mathfrak{R}^2}{\pi T}$$

The spectral energy density supplied when $N$ photons per second arrive over time a time interval of $T$ is:

$$U(\omega) = N \cdot P(\omega) \cdot T = \frac{NR|\Phi(\omega)|^2 \mathfrak{R}^2}{\pi}$$

This can be modeled as result:

$$\sigma_\Phi^2(\omega) = \frac{N|\Phi(\omega)|^2 \mathfrak{R}^2}{\pi} d\omega$$

$$\sigma_\Phi^2(f) = 2N|\Phi(\omega)|^2 \mathfrak{R}^2 df = 2N(hf)^2 \left(\frac{\eta q}{hf}\right)^2 df = 2N\eta^2 q^2 df$$

The probability of error for an *M*-ary DPSK using heterodyne detection has been derived in several texts. Following Park's [20] derivation for differential BPSK (henceforth referred to as DPSK) modulation by an optimal receiver is as follows:

The received signal is:

$$S_n(t) = \begin{Bmatrix} A\cos(\omega_c t + \theta) \text{ for } -T_b \leq t < 0 \\ A\cos(\omega_c t + \theta + n\pi) \text{ for } 0 \leq t < T_b \end{Bmatrix}$$

Where $n$ is 0 for should there be no change in the modulated data signal & 1 for a change, & $A\cos(\omega_c t + \theta) = x\cos(\omega_c t + \theta) + y\sin(\omega_c t + \theta)$. Given $\omega_c \gg \frac{2\pi}{T_b}$ he decision device computes:

$$l_n = (x_k x_{k-1} + y_k y_{k-1})\cos(n\pi) + (y_{k-1}x_k - y_k x_{k-1})\sin(n\pi)$$

$$l_n = (x_k x_{k-1} + y_k y_{k-1})\cos(n\pi)$$

Assuming no amplitude or incidental phase modulation between bits,

$$l_n = (x_k x_{k-1} + y_k y_{k-1}) \cdot \cos(n\pi)$$

$$l_0 = -l_1 = x_k x_{k-1} + y_k y_{k-1}$$

$$P_E = Prob(x_k x_{k-1} + y_k y_{k-1} < 0)$$

In the expression for $S_n$, if we set $\theta = 0$ & k = 1,

$$x_0 = \int_{-T_b}^{0} [A\cos(\omega_c t) + n(t)] \cdot \cos(\omega_c t)\, dt = \frac{A T_b}{2} + n_1$$

Note $n_1$ is a zero mean Gaussian random variable with $\sigma_{n_1}^2 = \frac{n_0 T_b}{4}$, $n_0$ is twice the one sided noise power spectral density.

$$x_1 = \frac{A T_b}{2} + n_2$$

$$y_0 = n_3$$

$$y_1 = n_4$$

Since $n_1, n_2, n_3$, & $n_4$ are zero mean Gaussian random variables with variances $\frac{n_0 T_b}{4}$,

$$P_E = Prob(x_k x_{k-1} + y_k y_{k-1} < 0)$$

$$P_E = Prob\left(\left(\frac{AT_b}{2} + n_1\right)\left(\frac{AT_b}{2} + n_2\right) + n_3 n_4 < 0\right)$$

Which can be rearranged,

$$P_E = Prob\left(\left(\frac{A T_b}{2} + \frac{n_1}{2} + \frac{n_2}{2}\right)^2 - \left(\frac{n_1}{2} - \frac{n_2}{2}\right)^2 + \left(\frac{n_3}{2} - \frac{n_4}{2}\right)^2 - \left(\frac{n_3}{2} - \frac{n_4}{2}\right)^2 < 0\right)$$

$$P_E = Prob\left(\left(\frac{A T_b}{2} + w_1\right)^2 - w_3^2 + w_2^2 - w_4^2 < 0\right)$$

$$P_E = Prob\left(\left(\frac{A T_b}{2} + w_1\right)^2 + w_2^2 < w_3^2 + w_4^2\right)$$

Where $w_1, w_2, w_3$, & $w_4$ are zero mean Gaussian random variables with variances $\frac{n_0 T_b}{8}$ since, for example $\overline{w_1^2} = \frac{\overline{n_1^2}}{4} + \frac{\overline{n_2^2}}{4}$. We have seen this equation before in the direct detection section, so we apply the results of that analysis:

$$P_E = \frac{1}{2} e^{-\frac{A^2 T_b}{2n_0}}$$

Because $\frac{A^2}{2}$ is the average value of a sinusoid of amplitude A squared, we take $\frac{A^2 T_b}{2}$ as the energy per bit, $E_b$, yielding the familiar result:

$$P_E = \frac{1}{2} e^{-\frac{E_b}{n_0}}$$

Modifying Gowar's [21] derivation of signal to noise ratio to be in terms of energies yields:

$$\frac{E_b}{n_0} = \frac{2\mathcal{R}^2 \Phi_R \Phi_L}{[(I_{sh}^*)^2 + (I_c^*)^2](\Delta f)}$$

Where $\mathcal{R}$ is the responsivity of the receiving photodetector, $\Phi_R$ is the received signal power, $\Phi_L$ is the power of the local laser, $e$ is the elementary charge, $I_{sh}^*$ is the mean shot noise per $\sqrt{Hz}$, $I_c^*$ is the mean current noise per $\sqrt{Hz}$ inducted by the amplifier, & $\Delta f$ is the bandwidth. Since $I_{sh}^* = 2 \cdot 2q\bar{I} = 2q\mathcal{R}(\Phi_R + \Phi_L)$, with $q$ is the elementary charge, if we assume the system is shot noise limited & the local laser power is much greater than the received power, we have:

$$\frac{E_b}{n_0} = \frac{2\mathcal{R}^2 \Phi_R \Phi_L}{[2q\mathcal{R}(\Phi_R + \Phi_L) + (I_c^*)^2](\Delta f)} \approx \frac{\mathcal{R}\Phi_R}{q(\Delta f)}$$

With $\mathcal{R}hf = e\eta$, $\eta$ the quantum efficiency & $q$ is the elementary charge,

$$\frac{E_b}{n_0} = \frac{\eta \Phi_R}{hf(\Delta f)}$$

If we assume the quantum efficiency is near 1 & $\Delta f$ the bandwidth, the $\frac{E_b}{n_0}$ is the number of photons per second $\left(\frac{\Phi_R}{hf} = \frac{received\ optical\ power}{power\ of\ one\ photon}\right)$ times the bit period $\left(\frac{1}{bandwidth}\right)$. Calling the number of photons per bit $N_{ph}$.

$$\frac{E_b}{n_0} = \eta N_{ph}$$

$$P_E = \frac{1}{2} e^{-\eta N_{ph}}$$